# A Random Matrix Approach to Wide Band Spectrum Sensing: Unknown Noise Variance Case

Sajjad Imani[1], Amin Banitalebi-Dehkordi[2], Mehdi Cheraghi[3]

[1] ECE. Dept., University of Tehran, Tehran, Iran, s.imani@ece.ut.ac.ir
[2] ECE. Dept., University of British Columbia, Vancouver, BC, Canada, dehkordi@ece.ubc.ca
[3] ECE. Dept., Shahed University, Tehran, Iran, cheraghi@shahed.ac.ir

*Abstract*—**In this paper three different scenarios in wide band spectrum sensing have been studied. While the signal and noise statistics are supposed to be unspecified, random matrixes have been utilized in order to estimate the noise variance. These scenarios are: 1- Number of subbands is specified and there is enough information regarding being used or being unused for each of them. 2- Number of subbands is known but there is no information about usage distribution among them. 3- Number of subbands is unknown. Simulation results showed the superior performance of the proposed scheme. Regarding the number of samples, the proposed method requires less number of samples compared to the cyclo-stationary spectrum sensing algorithms and more samples compared to the energy detection based methods. But, regarding the detection probability, the proposed method is superior compared to both other spectrum sensing methods.**

## I. INTRODUCTION

Cognitive Radio (CR) is known to be a system that can dynamically adjust its radio operating parameters to optimize the resource allocation in the system. Users of the system are divided to Primary/licensed Users (PU) and Secondary/unlicensed Users (SU). PUs own the authority of the spectrum while SUs transmit and receive signals over the licensed spectra or portions of it when the licensed users are inactive [1,2].

One of the key functionalities of CR is spectrum sensing which allows CRs to monitor and detect the unused spectrum in order to use the frequency spectrum optimally. Various spectrum sensing techniques are compared based on their accuracy and speed. Previous works have utilized various methods in order to sense the spectrum. Several techniques have been proposed to sense the spectrum. Techniques such as matched filtering, energy detection, cyclo-stationary feature detection and multi taper spectrum estimation [3]. Matched filtering maximizes the signal to noise ratio (SNR) which is not convenient for CR [4] as it requires knowledge about the PU's to demodulate the received signal and that's because a CR has minimum knowledge about the signal received. Spectrum sensing based on energy detection is most common method to detect the frequency holes or unused spectrum [5].

[6, 7] achieved proper spectrum sensing by linear combination of SU signal samples assuming that signal and channel parameters are known. Although these methods find the optimum combination coefficients, but they still suffer from inaccuracy problems. These non-blind approaches have not utilized combination of all signal samples in decision making. [8] used OFDM properties for spectrum sensing. In [9], wavelet transform has been utilized to find unused spectrum subbands.

In this paper we model the spectrum sensing into three different scenarios as follows: 1- Number of subbands in the frequency spectrum is specified and we know which subbands are being used and which are unused. 2- Number of subbands is known but there is no information about usage distribution for them. 3- Number of subbands is unknown. We assumed that there are free (unused) subbands in the spectrum which is reasonable in spectrum sensing application. In all of these three scenarios band width is supposed to be specified while Signal and noise statistics are assumed to be unknown. Random matrix theory is utilized in order to estimate the noise variance in three scenarios.

The rest of this paper is organized as follows: section II briefly reviews the random matrix theory. In section III we will introduce our noise variance estimation techniques in the three mentioned scenarios. Section IV contains spectrum sensing fidelity criteria. Section V illustrates the simulation results while section VI concludes the paper.

## II. OVERVIEW OF RANDOM MATRIX THEORY

Suppose $G$ is a $K \times N$ random matrix and all matrix elements have a zero-mean normal distribution with variance $\frac{\sigma^2}{N}$. When $K$ and $N$ become large numbers satisfying $\frac{K}{N} \to \varsigma$ then $GG^H$ will have a non-random distribution according to the Marcenko-Pastur law [10]:

$$f(t) = (1-\frac{1}{\varsigma})^+ u(t) + \frac{\sqrt{(t-a)^+(b-t)^+}}{2f\varsigma t}$$

$$x^+ = \begin{cases} x & x \geq 0 \\ 0 & x < 0 \end{cases} \quad (1)$$

In which *a* and *b* are defined as:

$$a = \dagger^2(1-\sqrt{s})^2$$
$$b = \dagger^2(1+\sqrt{s})^2 \qquad (2)$$

Note that *a* and *b* define the interval in which Eigen values of $GG^H$ can lie.

In order to utilize the random matrix theory in spectrum sensing we need to have a network upon which matrix *G* is constructed. We have used cooperative network with *K* users. Moreover, *N* samples are requested from each receiver and therefore *G* is built using these received samples. Figure 1 illustrates the configuration of the network.

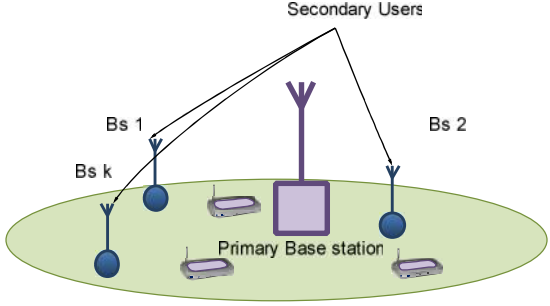

Figure 1. Cooperative network from which *G* is constructed

We assume that the receivers have the same noise distribution and noise variance should be estimated in case it is unknown.

## III. Noise Variance Estimation

### A. First Scenario

In the first scenario we assume that the number of the subbands is known and their usage distribution is a known distribution of *P(M)*. This is a reasonable assumption since the number of occurrences of a random event can be modeled with Erlang density function. Here we assume that the usage distribution in general follows *P(M)* distribution function and will estimate the noise based on this assumption.

Test hypothesis is as follows:

$$\underline{X} \sim \begin{cases} N(0,\dagger^2) & H_0 \\ N(0,\dagger_1^2) & H_1 \end{cases} \qquad (3)$$

$H_0$ is representing an empty channel while $H_1$ means that channel is being used by a primary user. Following this definition we have:

$$f(\underline{X}|H_0) = \frac{1}{(f\dagger^2)^M} e^{-\frac{1}{\dagger^2}\sum_k \|x_k\|^2} \prod_{k'} \frac{1}{f\dagger_{k'}^2} e^{-\frac{1}{\dagger_{k'}^2}\|x_{k'}\|^2} \qquad (4)$$

Where *k* is the total number of subbands and *M* is the number of unused subbands. Knowing the number of unused subbands, GLRT estimation gives the signal and noise estimation as [11]:

$$\dagger^2 = \frac{1}{ML}\sum_{k=1}^{M}\|x_k\|^2$$
$$\dagger_1^2 = \frac{1}{L}\sum_{k=1}^{M}\|X\|^2 \qquad (5)$$

Where *L* is the number of temporal samples. Substituting these into (4) we have:

$$f(\underline{X}|H_0) = \frac{1}{f^M(\frac{1}{ML}\sum_{k=1}^{M}\|x_k\|^2)^M} e^{-\frac{1}{\frac{1}{ML}\sum_{k=1}^{M}\|x_k\|^2}\sum_k \|x_k\|^2} \times \prod_{k'}\frac{1}{f\dagger_1^2} e^{-\frac{1}{\dagger_1^2}\|x_{k'}\|^2} \qquad (6)$$

$$f(\underline{X}|H_0) = \frac{1}{f^M(\frac{1}{ML}\sum_{k=1}^{M}\|x_k\|^2)^M} e^{-ML} \times \prod_{r=M+1}^{k}\frac{1}{f(\frac{1}{L}\|X\|^2)} e^{-L}$$

Then $f(M)$ is defined by:

$$f(M) = \frac{1}{f^M(\frac{1}{ML}\sum_{k=1}^{M}\|x_k\|^2)^M} e^{-ML} \times \prod_{r=M+1}^{k}\frac{1}{f(\frac{1}{L}\|X\|^2)} \qquad (7)$$

Finally, an *M* which maximizes equation (7) can be derived from:

$$\hat{M} = \arg\min\{M\log(\frac{1}{M}\sum_{k=1}^{M}\|\underline{X}_k\|^2) + \sum_{r=M+1}^{k}\log(\|\underline{X}_r\|^2) - \log(P(M))\} \qquad (8)$$

### B. Second Scenario

In the second case, number of subbands is known but there is no information about usage distribution for them. Following the similar steps as discussed about the previous scenario, we have:

$$\hat{M} = \arg\min\{M\log(\frac{1}{M}\sum_{k=1}^{M}\|\underline{X}_k\|^2) + \sum_{r=M+1}^{k}\log(\|\underline{X}_r\|^2) - \log(P(M))\} \qquad (9)$$

This solution is the same as solution provided in [11]. Since number of subbands is specified, therefore noise variance can be claimed to be equal to energy of the subband with minimum energy among all subbands.

### C. Third Scenario

In the third scenario, number of subbands is unknown. We assume that there are free subbands among the whole set of subbands and width of all subbands is equal to the subband with minimum width. In order to find the subband with minimum width, we follow a similar approach as [12-14].

Next, number of subbands, *k*, can be found by dividing total band width by width of the each subband. Finally, whether usage distribution of the subbands is specified or not, one can use equation (8) or (9) to estimate the noise variance.

## IV. Spectrum Sensing

Consider figure 1. In this configuration, primary users are communicating with base stations. Secondary users are sensing the spectrum cooperatively in order to find the unused bands. Suppose there are *K* secondary users wired connected to a secondary base station which processes the information to make final decisions. Each user receives $N$ samples, therefore:

$$Y = \frac{1}{\sqrt{N}} \begin{pmatrix} y_1(1) & \cdots & y_1(N) \\ \vdots & \ddots & \vdots \\ y_k(1) & \cdots & y_k(N) \end{pmatrix} \quad (10)$$

$Y$ can be rewritten as:

$$Y = \frac{1}{\sqrt{N}} \begin{pmatrix} h_1 & \sigma & & 0 \\ \vdots & & \ddots & \\ h_k & 0 & & \sigma \end{pmatrix} \begin{pmatrix} s(1) & s(2) & \cdots & s(N) \\ z_1(1) & z_1(2) & \cdots & z_1(N) \\ \vdots & \vdots & & \vdots \\ z_K(1) & z_K(2) & \cdots & z_K(N) \end{pmatrix} \quad (11)$$

Suppose signal samples are independent from each other and from noise samples. Also, noise samples have independent distributions. $G$ can be written as:

$$G = \begin{pmatrix} h_1 & \sigma & & 0 \\ \vdots & & \ddots & \\ h_k & 0 & & \sigma \end{pmatrix} \quad (12)$$

Then Eigen values of matrix $GG^H$ are in the form of:

$$\lambda_j = \begin{cases} \sum_i |h_i|^2 + \sigma^2 & j = 1 \\ \sigma^2 & j \neq 1 \end{cases} \quad (13)$$

Marcenko-Pastur law assumption is not satisfied in the current case since neither the number of secondary users nor the number of samples, go to infinity (having few number of samples is a feature of high performance spectrum sensing mechanisms). As a result, Eigen values are distributed around the noise variance which was found in section III. In [15] spectrum sensing is based on the ratio of geometric mean over the algebraic mean of the Eigen values of the received matrix. However, this method has acceptable performance only the low signal to noise ratios (SNR). Therefore, a metric has to be defined as a function of Eigen values. Then, decision making is done by setting thresholds for the specified metric. Choosing a threshold value which is greater than 1 reduces the detection probability ($P_D$) while a threshold value less than unity results in an increase in false alarm probability ($P_{FA}$).

We propose a new threshold scheme based on the most effective Eigen values as follows:

$$T = \frac{\lambda_{\max}}{C} = \frac{\sum_i |h_i|^2 + \sigma^2}{C} > \eta \quad (14)$$

Where, $C$ is the biggest Eigen value. Monte Carlo methods are utilized to find the most appropriate threshold value. Also, for $N$ values greater than 50 and for $K$ values greater than 5 ($\sigma < 5$) [16], values of $\lambda_{\max}, C$ are found from the diagram of equation (1).

## V. EXPERIMENTAL RESULTS

Figure 2.a shows a typical normalized spread spectrum signal. In this example, SNR ratio is -5 dB. Suppose there is no information about the number of channels and $P(M)$ distribution. The goal is to detect the unused subbands. Figure 2.b demonstrates frequency holes detection based on our previous work, described in [16]. In [12], we proposed to consider a subband between each pair of neighbor local maximums. Figure 3 shows the noise estimation error when frequency band is divided into equal-width subbands. As it can be seen from this figure, small number of sub channels will result in big error values, i.e. estimation is more accurate when number of channels is high enough. But increasing the number of subbands will increase the computational complexity. Figure 4 demonstrates the estimation error when the frequency spectrum is divided to subbands based on the algorithm proposed in [12]. A typical subbands distribution can be seen in figure 2.a. As we see, noise estimation error is less than 5 percent.

As we discussed before, appropriate threshold value is found using Monte Carlo methods. Figure 5 sketches the false alarm probability versus the threshold value. Appropriate linear estimation can be used to estimate this diagram in order to solve the spectrum sensing problem in closed form. We used 7 receivers in the simulation to set up the cooperative network, each receiving 100 samples ($K=7, N=100$). Figure 6 shows the detection probability in terms of the false alarm probability when $SNR=-10$ dB. The proposed algorithm in [15] is sensitive to the threshold value. This lowers the performance in small false alarm probabilities (Figure 6). Figure 7 verifies the high performance of our proposed method even in small SNR values. Regarding the number of samples, our method requires less number of samples compared to the cyclo-stationary methods on average, which reduces the complexity. Comparing with energy detection based algorithms, number of samples is higher, but the performance is significantly better in small SNR values. Therefore it becomes as a trade-off between performance and complexity.

## VI. CONCLUSION

In this paper we utilized our previously proposed frequency-hole detection mechanism and a new noise estimation method to improve the spectrum sensing performance. We studied the problem in three different scenarios. When signal and noise statistics are supposed to be unspecified, random matrixes have been utilized in order to estimate the noise variance. We proposed new threshold detection metric based on the Eigen values of the constructed random matrix. Simulation results showed the superior performance of the proposed scheme. Regarding the number of samples, the proposed method requires less number of samples compared to the cyclo-stationary spectrum sensing algorithms and more samples compared to the energy detection based methods. But, regarding the detection probability, the proposed method is superior compared to both other spectrum sensing methods.

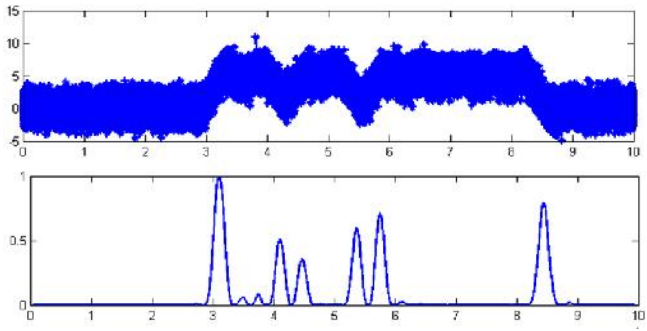

Figure 2. a) Spread spectrum wideband signal b) Frequency holes detection with approach stated in [14]

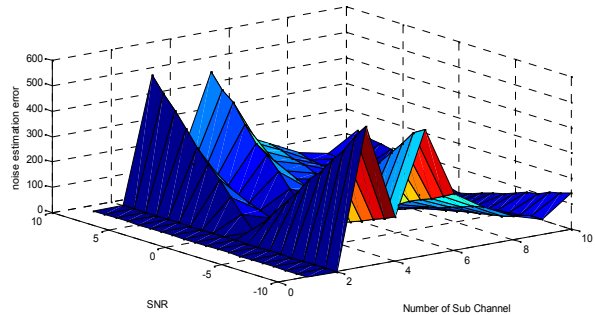

Figure 3. Noise estimation error when dividing the frequency band into equal subbands

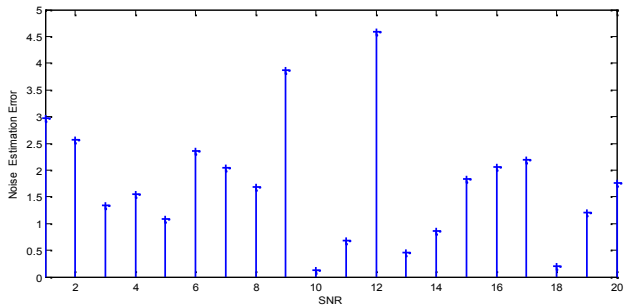

Figure 4. Noise estimation error when dividing the frequency band into subbands based on the algorithm proposed in

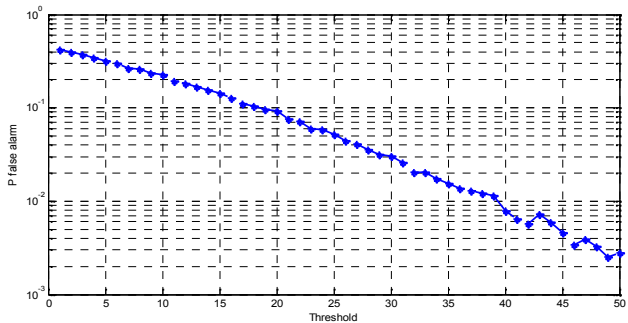

Figure 5. False alarm probability versus detection threshold

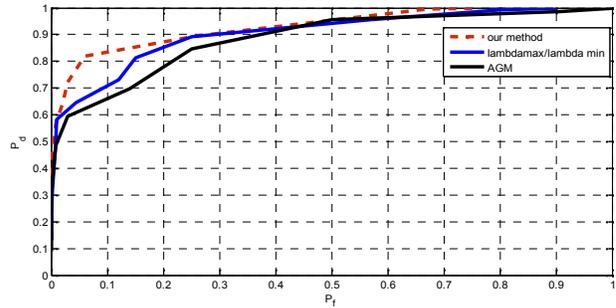

Figure 6. Detection probability versus false detection probability

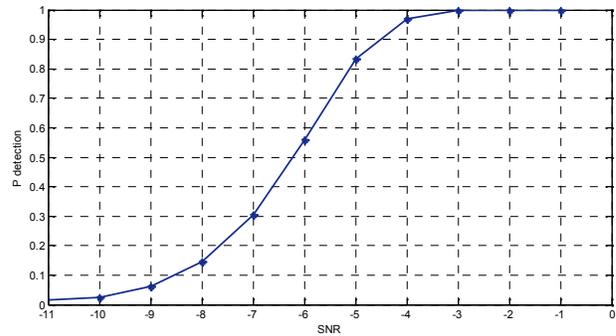

Figure 7. Detection probability versus SNR